\documentclass[preprint]{aastex}

\shortauthors{Burgasser et al.}
\shorttitle{2MASS J0559-14}

\begin{document}

\title{Discovery of a Bright Field Methane (T-type) Brown Dwarf by 2MASS}

\author{
Adam J.\ Burgasser\altaffilmark{1},
John C.\ Wilson\altaffilmark{2},
J.\ Davy Kirkpatrick\altaffilmark{3},
Michael F.\ Skrutskie\altaffilmark{4},
Michael R.\ Colonno\altaffilmark{2},
Alan T.\ Enos\altaffilmark{2},
J.\ D.\ Smith\altaffilmark{2},
Charles P.\ Henderson\altaffilmark{2},
John E.\ Gizis\altaffilmark{3}, 
Michael E.\ Brown\altaffilmark{5,6},
and
James R.\ Houck\altaffilmark{2}
}
 
\altaffiltext{1}{Division of Physics, M/S 103-33, 
California Institute of Technology, Pasadena, CA 91125; diver@its.caltech.edu}
\altaffiltext{2}{Space Sciences, Cornell University, Ithaca, NY 14853;
jcw14@cornell.edu, mrc10@cornell.edu, ate1@cornell.edu, 
jdsmith@astrosun.tn.cornell.edu, cph5@cornell.edu, jrh13@cornell.edu}
\altaffiltext{3}{Infrared Processing and Analysis Center, M/S 100-22, 
California Institute of Technology, Pasadena, CA 91125; davy@ipac.caltech.edu,
gizis@ipac.caltech.edu}
\altaffiltext{4}{Five College Astronomy Department, Department of Physics
and Astronomy, University of Massachusetts, Amherst, MA 01003;
skrutski@north.astro.umass.edu}
\altaffiltext{5}{Division of Geological and Planetary Sciences, M/S 105-21, 
California Institute of Technology, Pasadena, California 91125;
mbrown@gps.caltech.edu}
\altaffiltext{6}{Alfred P.\ Sloan Research Fellow}

\begin{abstract}
We report the discovery of a bright (J = 13.83$\pm$0.03) methane
brown dwarf, or T dwarf, by the Two Micron All Sky Survey.
This object, 2MASSI J0559191-140448, is the first brown dwarf
identified by the newly commissioned CorMASS instrument mounted on
the Palomar 60-inch Telescope.  Near-infrared spectra from 0.9 - 2.35 $\micron$
show characteristic CH$_4$ bands at 1.1, 1.3, 1.6, and 2.2 $\micron$,
which are significantly
shallower than those seen in other T dwarfs discovered to date.  
Coupled with the detection of an FeH band at 0.9896 and
two sets of K I doublets at J-band, 
we propose that 2MASS J0559-14 is a warm T dwarf,
close to the transition between L and T spectral classes.  The
brightness of this object makes it 
a good candidate for detailed investigation over a 
broad wavelength regime and at higher resolution.
\end{abstract}


\section{Introduction}

T dwarfs are brown dwarfs that exhibit methane absorption bands at 
1.6 and 2.2 $\micron$ \citep{Ki99a}, and thus have
effective temperatures T$_{eff}$ $\lesssim$ 1200-1300 K
\citep{Fe96,Bu99,Ki00}.  The prototype for this class, Gl 229B \citep{Na95,Op99}, was
identified as a cool companion to the nearby M1V star Gl 229A.  
Recently, seven field objects \citep{Ss99,Bg99,Cu99,Ts00} and
another companion object \citep{Bg00a} have also been identified as T dwarfs.
The rapid discovery of these cool brown dwarfs has been driven by new
sky surveys, such as the Two Micron All Sky Survey
\citep[hereafter 2MASS]{Sk97} and the Sloan Digital Sky Survey \citep{Gu95};
and deep near-infrared surveys, such as the ESO
New Technology Telescope Deep Field \citep{Ar99}. 

The T dwarfs identified to date are remarkably similar to Gl 229B, with 
colors in the range $-$0.2 $\lesssim$ J-K$_s$ $\lesssim$ 0.2.  
Near-infrared spectra are correspondingly similar
\citep{Ss99,Bg00a},
likely due to the saturation of H$_2$O and CH$_4$
bands that dominate this wavelength regime. 
Subtle differences in the magnitudes and shapes of
H- and K-band flux peaks are observed,
due to increased CH$_4$, H$_2$O, and H$_2$ collision-induced absorption (CIA)
toward cooler effective temperatures \citep{Bg99,Ts00},
and variations in the depths of near-infrared H$_2$O and CH$_4$ bands are
discerned when compared to Gl 229B \citep{Na00,Bg00a}.
Nonetheless, the similarity of the near-infrared spectra 
suggests that either the objects thus far identified are very
similar in temperature, around 1000 K \citep{Ma96}, or that near-infrared
features are fairly insensitive to temperature, making the definition of a
T dwarf spectral sequence in this wavelength regime
a difficult proposition, at least at
low resolution.

We report the discovery of a T dwarf by 2MASS which is unique
among its counterparts, as it shows significant differences in its near-infrared
features while retaining defining CH$_4$ bands.  This object,
2MASSI J0559191-140448 (hereafter 2MASS J0559-14), is also 0.4 mag brighter
than Gl 229B and more than 1 mag brighter than the field T dwarfs discovered 
thus far.  It is the first brown
dwarf to be identified by the newly commissioned 
Cornell Massachusetts Slit Spectrograph
\citep[hereafter CorMASS]{Wi00}, mounted on the Palomar 60-inch
telescope.  In $\S$2 we discuss the selection of 2MASS J0559-14 from
2MASS data and its spectral identification by CorMASS.  In $\S$3 we
discuss the observed spectral features and argue that 2MASS J0559-14 is
a warm T dwarf, possibly close to the transition
temperature between L and T spectral classes.  
We discuss the brightness of this object and its role in future
spectroscopic investigation of the T dwarf class in $\S$4.

\section{Identification of 2MASS J0559-14}

\subsection{Selection and Confirmation}

2MASS J0559-14 was initially selected as a T dwarf candidate 
from the 2MASS point-source working database.  
Details on the selection criteria for
2MASS T dwarfs are discussed in \citet{Bg00a}.
Using the
Palomar 60-inch Infrared Camera \citep{Mu95}, we successfully
re-imaged this candidate
on 23 September 1999 (UT) at J-band, confirming it
as a bona-fide candidate (i.e., not a
minor planet or artifact).
2MASS J0559-14 is the brightest 2MASS T dwarf candidate 
confirmed to date, with J = 13.83$\pm$0.03 (Table 1); it is
also the reddest candidate confirmed, with J - K$_s$ = 0.22$\pm$0.06.
Optical SERC-EJ \citep{Mr92}
and J-band 2MASS images of the 
2MASS J0559-14 field are shown in Figure 1; no optical counterpart is seen
to B$_J$ $\sim$ 23.  

\placefigure{fig-1}
\placetable{tbl-1}

\subsection{Spectral Identification with CorMASS}

Low-resolution near-infrared spectral data of 2MASS J0559-14 were obtained 
on 24 October 1999 (UT) using CorMASS.
This newly commissioned instrument 
is an R $\sim$ 300 prism cross-dispersed near-infrared
spectrograph, with a 256 $\times$ 256 NICMOS3 array and a 40 lines 
mm$^{-1}$ grating, blazed at 4.8 $\micron$. CorMASS was designed
primarily for the spectral classification of candidate low-mass
objects color-selected from the 2MASS database. The instrument's
echelle format provides simultaneous coverage of the {\it z\/}JHK
bands, for 0.8 $\lesssim$ $\lambda$ $\lesssim$ 2.5
$\micron$. The fixed slit has a width of 2$\arcsec$ and a length
of 15$\arcsec$.  Further details on this instrument can be
found in \citet{Wi00}.

Conditions during the observations were not photometric, and estimated
seeing was $\sim$ 1$\farcs$5, with variable thin cirrus throughout
the night.  Total on-source time was 2400 seconds divided into sets of
300-second integrations, nodding $\sim$ 5$\arcsec$ along the slit between
exposures.
Spectra were reduced using standard IRAF routines.  After correction
for bad pixels, flat field and flux calibration images were corrected
for NICMOS3 reset-decay bias (also known as shading) as follows: a
quadratic was fit to a row-by-row clipped median of the top quarter of
the array, the portion unused by the spectrograph.  The fit was
extrapolated to 128 rows for the top two quadrants, duplicated for the
bottom two quadrants, and subtracted row-by-row from all pixels.
All images were flat-fielded with a pixel responsivity solution from
the APFLATTEN task using dome flats from all five nights of the run
summed together.  Nodded image pairs for the science object were
subtracted against each other to remove sky background and reset-decay
bias.  The APALL task was used for spectral extraction.  Wavelength
calibration was accomplished using spectral observations of the
planetary nebula NGC 7027 obtained on 25 October 1999 (UT).  Flux
calibration was done by dividing by the reduced spectrum of the A2V
standard HD 77281 \citep{El82}, which was observed at
similar airmass as 2MASS J0559-14 and hand corrected for stellar H
Paschen and Brackett recombination lines.  The ratio was then
multiplied by a 8810 K \citep{To00} blackbody to complete the flux calibration.
Finally, all 300-second observations for each order
were weighted by the spectra mean and
combined with the SCOMBINE task using a trimmed average.  
Orders were stitched
together with SCOMBINE after hand deletion of noisy data at the
ends of the orders.

The spectrum of 2MASS J0559-14 is shown in Figure 2, along with 
optical (0.87-1.0 $\micron$; Burgasser et al.\ 2000b) 
and near-infrared (1.0-2.35 $\micron$; Strauss et al.\ 1999) data 
for SDSSp J162414.37+002915.6 (hereafter SDSS 1624+00).  
The combined SDSS 1624+00
spectrum was smoothed to the resolution of CorMASS.
Spectra for both objects
are normalized to one at 1.27 $\micron$ (J-band peak).  Close up
views of 0.88-1.01 and 1.15-1.345 $\micron$
are shown in Figure 3.

\section{Discussion of the Near-Infrared Spectrum}

\subsection{Spectral Features}

Table 2 summarizes the spectral features 
detected in 2MASS J0559-14,
with identifications from
\citet{Pg63}, \citet{Da66}, \citet{Ws66}, and \citet{Ph87}.
Only major absorption
bands of H$_2$O and CH$_4$ are tabulated.  The characteristic CH$_4$ bands
at 1.6 and 2.2 $\micron$ are present, as are bands at 1.1 and 1.3 $\micron$
identified from laboratory data \citep{Fi79} which are blended with 
H$_2$O bands at the same wavelengths.  
An FeH feature is seen at 0.9896 $\micron$ 
(0-0 band of A$^4$$\Delta$-X$^4$$\Delta$) which has also
been identified in SDSS 1624+00 \citep{Bg00b}.  
We do not detect the higher order 0-1 FeH band at 1.191
$\micron$, which is seen to weaken in the latest L dwarfs \citep{Mc00}.
Two sets of K I doublets are noted at
1.1690 \& 1.1773 $\micron$ (4p $^2$P$_0$ $-$ 3d $^2$D) and 1.2432 \& 1.2522 
$\micron$ (4p $^2$P$_0$ $-$ 5s $^2$S), as is Cs I 
at 0.8943 $\micron$ (6s $^2$S$_{1/2}$ $-$ 6p $^2$P$_{1/2}$).
We do not detect the 2-0 X$^1$$\Sigma$$^+$-X$^1$$\Sigma$$^+$ band of CO
at 2.3 $\micron$.

Comparison between 2MASS J0559-14 and SDSS 1624+00 reveals significant
differences in spectral morphology.  
In Figure 2, it is apparent that the slope between 0.9 and 1.05 $\micron$ 
is shallower in 2MASS J0559-14, likely 
due to decreased absorption by the pressure-broadened K I doublet at 0.7665
and 0.7699 $\micron$ \citep{Li00}.
CH$_4$ and H$_2$O features are generally weaker in 
2MASS J0559-14, as noted by the significantly
weakened 1.1 - 1.2 $\micron$ trough between $z$- and J-band peaks.
The decreased CH$_4$ opacity noticeably affects the shape of
the J-band peak near 1.27 $\micron$, as the weak CH$_4$
wings at 1.24 and 1.30 $\micron$ carve out less
flux on either side of the peak.  
In Figure 2, a significant flux offset is
readily apparent at the base of the 
1.6 $\micron$ CH$_4$ band in 2MASS J0559-14, and
relative enhancement of flux at both
H- and K-bands in this object
is almost certainly due to decreased H$_2$ CIA (1-0 Quadrupole), 
H$_2$O, and CH$_4$ opacity.  

\placefigure{fig-2}
\placefigure{fig-3}
\placetable{tbl-2}

\subsection{2MASS J0559-14 is a Warm T dwarf}

The weak CH$_4$ bands seen in 2MASS J0559-14 are unique among the current sample
of T dwarfs, and the reduced opacity can be most readily explained if this object
is warmer than other known T dwarfs.  At higher effective temperatures,
the dominant carbon-bearing species changes from CO to CH$_4$ 
at small optical depth, so that the
CH$_4$ column density will be less than that of cooler T dwarfs,
and observed band strengths correspondingly weaker. 
Increased thermal flux will also be seen at the base of these bands, 
particularly at 1.6 $\micron$, which is unaffected by H$_2$O absorption.  
The lower CH$_4$ column density directly affects the 
H$_2$O column density, via the reaction
CO + 3 H$_2$ $\rightarrow$ CH$_4$ + H$_2$O, leading to
shallower bands at 1.1 and 1.45 $\micron$.
Water can also be heated and dissociated by dust layers deep in 
the photosphere \citep{Le98}.
Finally, decreased H$_2$ CIA opacity at K-band, congruous 
with reduced CH$_4$ and H$_2$O opacity, will result in redder J-K$_s$ colors
with warmer T$_{eff}$.
These features are observed in 2MASS J0559-14, and its
warm temperature is independently supported by the
detection of the 0.9896 $\micron$ FeH band, which is present but weakening
in the latest L dwarfs \citep{Ki99a}.  A similar argument has been made
for SDSS 1624+00 by \citet{Bg00b}, which was one of only three T dwarfs in
that paper to show this feature.
SDSS 1624+00 has been shown to be a warm object via optical continuum
measurements
between broadened Na I and K I features \citep{Li00}, while \citet{Na00}
argue that this object is both warmer and dustier than Gl 229B based on
its shallower CH$_4$ and H$_2$O bands.
By analogy, 2MASS J0559-14 should be warmer still.  The detection of
excited K I lines at J-band, which are seen to weaken in the latest
L dwarfs \citep{Mc00}, is further evidence of the warmth of this object.

The spectral features in 2MASS J0559-14 suggest that
it is close to the L/T transition temperature.  
The lack of CO detection at
2.3 $\micron$ is not necessarily contradictory to this hypothesis, 
as overlying CH$_4$ absorption beyond 2.2 $\micron$
may mask this weaker feature.   
Naturally, metallicity, dust, and gravity could also play roles in the 
band strengths seen in 2MASS J0559-14; however, temperature is likely to be the
dominant determinant given the concordance of spectral features as discussed
above.  
We can make a conservative T$_{eff}$ constraint for this object
based on the temperature of Gl 229B, which is clearly cooler, and 
a temperature estimate of the L8V companion dwarf Gl 584C \citep{Ki00}; 
this translates
into a range of 1000 $\lesssim$ T$_{eff}$ $\lesssim$ 1300 K.
Parallax and bolometric luminosity measurements of this object 
would allow a direct determination of temperature.

\section{The Brightness of 2MASS J0559-14}

The relative brightness of 2MASS J0559-14 as compared to other T dwarfs
suggests that it may be a nearby brown dwarf.  
We can estimate its distance using simple scaling arguments based on
Gl 229B\footnote{In this section we adopt the following values for Gl 229B:
$d$ = 5.77$\pm$0.04 pc \citep{Pe97}, 
T$_{eff}$ = 960$\pm$70 K \citep{Ma96}, 
R $\approx$ R$_{Jup}$ = 7.1 $\times$ 10$^9$ cm \citep{Bu93},
J = 14.32$\pm$0.05,
L = (6.6$\pm$0.6) $\times$ 10$^{-6}$ L$_{\sun}$,
and BC$_J$ = 2.2$\pm$0.1 \citep{Le99}.}.  
First, if we assume this object has 
the same intrinsic luminosity as Gl 229B, we estimate its distance to be 
4.6 pc.
However, 2MASS J0559-14 is probably a warmer object than
Gl 229B, so it should be more distant.  
An alternate estimate can be made if we adopt a radius
R $\approx$ R$_{Gl229B}$ and bounding T$_{eff}$ $\lesssim$ 1300 K, 
so that L
$\lesssim$ 2.2 $\times$ 10$^{-5}$ L$_{\sun}$.  
Using a Gl 229B bolometric correction,
we obtain M$_J$ $\gtrsim$ 14.2 and thus $d$ $\lesssim$ 8.4 pc.
Note, however, that an L8V has M$_J$ = 15.0 \citep{Ki00}, implying 
that our BC$_J$ value may be too high; indeed, BC$_J$ 
may be only 1.3 for an L8V \citep{Rd99}.  
Adopting a smaller BC$_J$ yields
a smaller distance.  
Therefore, barring multiplicity, 2MASS J0559-14
is probably 5-8 pc distant, and likely falls
within the 8 pc nearby star sample defined by \citet{Rd97}.

The brightness of this object allows substantial follow-up over a broad 
wavelength range, particularly for $\lambda$ $>$ 2.5 $\micron$, where
the fundamental absorption bands of H$_2$O
(2.7 $\micron$), NH$_3$ (3.0 $\micron$),
CH$_4$ (3.3 $\micron$), H$_2$S (3.8 $\micron$),
and CO (4.7 $\micron$) are found.  The latter
1-0 X$^1$$\Sigma$$^+$-X$^1$$\Sigma$$^+$ band of CO 
can aid in constraining its atmospheric abundance \citep{No97},
essential in testing hypotheses of mixing \citep{Op98,Gr99} and 
transparency \citep{Lo99} in T dwarf atmospheres.  Deuterated molecules,
such as CH$_3$D, are of interest for study in the 3-5 $\micron$ range,
as they yield information on the Deuterium burning history of these
brown dwarfs.
Mid-infrared wavelengths
(8 - 14 $\micron$) are sensitive to vibration-rotation bands 
of alkali chlorides and sulfides, which are more abundant than their atomic
alkali counterparts 
at low temperatures and high pressures \citep{Lo99} and may serve
as excellent temperature discriminants.  This region also contains the 
strong fundamental $\nu$$_2$
band of NH$_3$ (10.5 $\micron$) and various silicate features.  

Additionally, optical spectra shortward of 
0.9 $\micron$ are critical in determining the behavior of the alkalis in cool
brown dwarfs, particularly pressure-broadened Na I (0.5890 \& 0.5896 $\micron$)
and K I (0.7665 \& 0.7699 $\micron$) doublets, and Cs I lines at 0.8521 and
0.8943 $\micron$, which may also be used
as temperature discriminants \citep{Rd00,Bu00,Ba00}.  While these 
features have been
seen in T dwarfs such as SDSS 1624+00 \citep{Li00}, 
detailed investigation of Li I (0.6708 $\micron$)
and Rb I (0.7800 \& 0.7948 $\micron$) 
lines have been hampered by the lack of detectable flux at these
wavelengths.  
Investigation in this spectral
regime is currently underway.

\acknowledgements

We thank M.\ Strauss for the kind 
use of the SDSS 1624+00 near-infrared spectrum and 
acknowledge useful discussions with C.\ Griffith and M.\ Marley.
We thank our anonymous referee for helpful comments and suggestions.
We also recognize the efforts of the 2MASS staff and scientists 
in creating a truly incredible astronomical resource, and the expert 
assistance
of the Palomar Observatory staff during imaging and spectroscopic 
observations.
A.\ J.\ B., J.\ D.\ K., and J.\ E.\ G.\ acknowledge the support 
of the Jet Propulsion
Laboratory, California Institute of Technology, which is operated under
contract with the National Aeronautics and Space Administration.  
J.\ C.\ W.\ acknowledges support by NASA grant NAG5-4376. 
Observations from Palomar Observatory
were made as part of a continuing collaboration between the
California Institute of Technology and Cornell University.
The Digitized Sky Survey was produced at the Space
Telescope Science Institute under US Government grant NAG W-2166.
SERC-EJ data were scanned by DSS from photographic data obtained using 
the UK Schmidt Telescope, operated by the Royal Observatory
Edinburgh, with funding from the UK Science and Engineering Research 
Council.  
DSS images were obtained from the Canadian Astronomy Data Centre, 
which is operated by the Herzberg Institute of Astrophysics, 
National Research Council of Canada. 
IRAF is distributed by the
National Optical Astronomy Observatories, which are operated by
the Association of Universities for Research in Astronomy, Inc.,
under cooperative agreement with the National Science
Foundation.
This publication makes use of data from the Two Micron
All Sky Survey, which is a joint project of the University of
Massachusetts and the Infrared Processing and Analysis Center,
funded by the National Aeronautics and Space Administration and
the National Science Foundation.

\clearpage

\figcaption[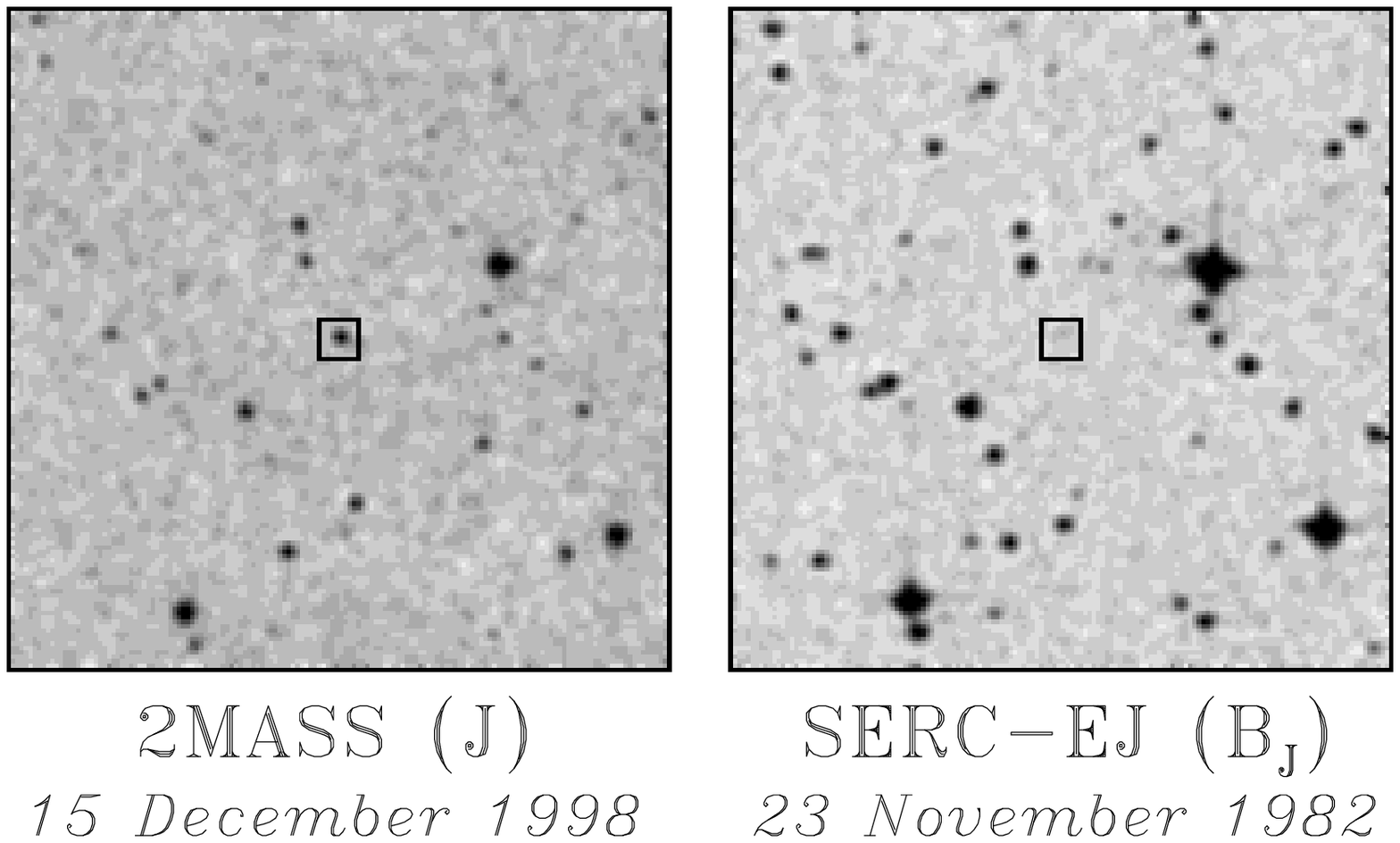]{SERC-EJ Optical and 2MASS J-band images of the
2MASS J0559-14 field.  Images are 5$\arcmin$ $\times$ 5$\arcmin$
with North up and East to the left.  A
20$\arcsec$ $\times$ 20$\arcsec$ box 
is drawn around the location of the T dwarf
in both images.
\label{fig-1}}

\figcaption[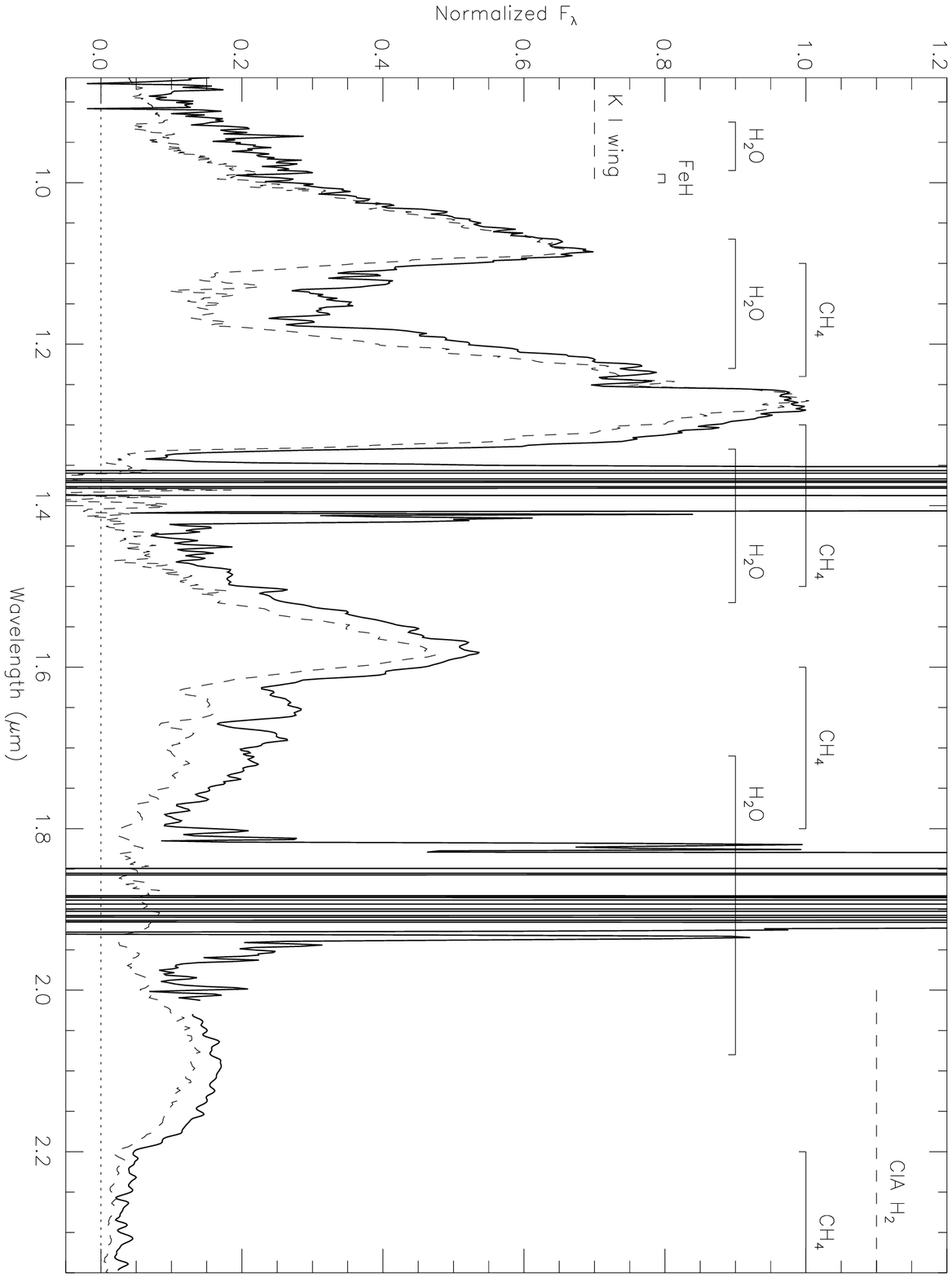]{Near-infrared spectrum of 2MASS J0559-14 (solid line)
obtained by
CorMASS.  Overlaid are SDSS 1624+00 data from \citet{Bg00b} and from \citet{Ss99}
for 0.87 - 1 $\micron$ and 1 - 2.35 $\micron$, respectively
(dashed line).  Both spectra are normalized to one at the 1.27 $\micron$ peak. 
Prominent molecular bands of H$_2$O, CH$_4$, and H$_2$ (collision-induced
absorption) are indicated, as is an FeH band at 0.9896 $\micron$
and the K I wing at $z$-band. 
\label{fig-2}}

\figcaption[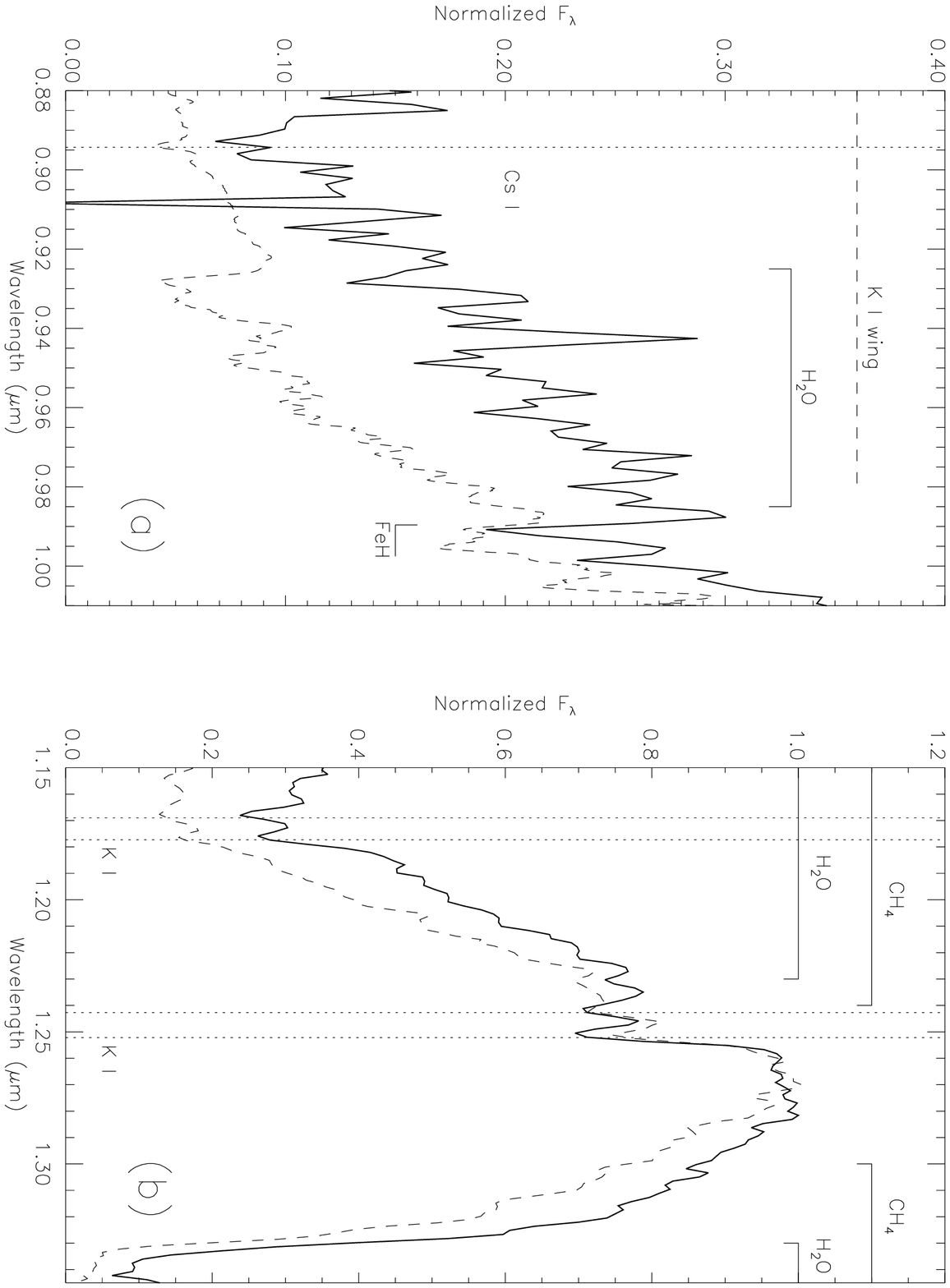]{Two detailed regions of Figure 2 showing
(a) 0.88 - 1.01 $\micron$ and (b) 1.15 - 1.345 $\micron$.
Spectra are normalized as in Figure 2.
Features of FeH (0.9896 $\micron$),
K I (1.1690, 1.1773, 1.2432, 1.2522 $\micron$), and Cs I (0.8943 $\micron$)
are indicated, as are H$_2$O and CH$_4$ bands.
\label{fig-3}}

\clearpage

\begin{deluxetable}{ccccccc}
\tabletypesize{\scriptsize}
\tablenum{1}
\tablewidth{0pt}
\tablecaption{Photometric Properties of 2MASSI J0559191-140448\tablenotemark{a}. \label{tbl-1}}

\tablehead{
\colhead{J} &
\colhead{H} &
\colhead{K$_s$} &
\colhead{J - H} &
\colhead{H - K$_s$} &
\colhead{J - K$_s$} &
\colhead{Estimated Distance (pc)}
}
\startdata
13.83$\pm$0.03 & 13.68$\pm$0.04 & 13.61$\pm$0.05 & 0.15$\pm$0.05 & 0.07$\pm$0.06 & 0.22$\pm$0.06 & 5 - 8\tablenotemark{b} \\
 
\tablenotetext{a}{Source designations for 2MASS sources in the Incremental
Release Catalogs are given
as ``2MASSI Jhhmmss[.]s$\pm$ddmmss''.  The suffix conforms to IAU
nomenclature convention and is the sexigesimal R.A. and decl. at J2000 equinox.} &
\tablenotetext{b}{See $\S$4 for discussion.}
\enddata
\end{deluxetable}

\clearpage

\begin{deluxetable}{cccl}
\tabletypesize{\scriptsize}
\tablenum{2}
\tablewidth{0pt}
\tablecaption{Near-Infrared Features Detected in 2MASS J0559-14.\label{tbl-2}}

\tablehead{
\colhead{Feature} &
\colhead{$\lambda$ ($\micron$)} & 
\colhead{Transition}  &
\colhead{Reference for Transition}
}

\startdata
Cs I & 0.8943 & 6s $^2$S$_{1/2}$ $-$ 6p $^2$P$_{1/2}$ & \citet{Ws66} \\
K I & broadened up to $\sim$ 0.9 & 4s $^2$S$_{1/2}$ $-$ 4p $^2$P$_{3/2,1/2}$  & \citet{Bu00} \\
H$_2$O & 0.925 - 0.95 & $\nu$$_2$ = 0, $\nu$$_1$ + $\nu$$_3$ = 3 & \citet{Au67} \\
H$_2$O & 0.95 - 0.985 & $\nu$$_2$ = 2, $\nu$$_1$ + $\nu$$_3$ = 2 & \citet{Au67} \\
FeH & 0.9896 & 0-0 band of A$^4$$\Delta$-X$^4$$\Delta$ & \citet{Ph87} \\
H$_2$O & 1.07 - 1.11 & $\nu$$_2$ = -1, $\nu$$_1$ + $\nu$$_3$ = 3 & \citet{Au67} \\
CH$_4$ & 1.1 - 1.24 & 3$\nu$$_3$ & \citet{Da66} \\
H$_2$O & 1.11 - 1.16 & $\nu$$_2$ = 1, $\nu$$_1$ + $\nu$$_3$ = 2 & \citet{Au67} \\
H$_2$O & 1.16 - 1.23 & $\nu$$_2$ = 3, $\nu$$_1$ + $\nu$$_3$ = 1 & \citet{Au67} \\
K I & 1.1690 & 4p $^2$P$_0$ $-$ 3d $^2$D & \citet{Ws66} \\
K I & 1.1773 & 4p $^2$P$_0$ $-$ 3d $^2$D & \citet{Ws66} \\
K I & 1.2432 & 4p $^2$P$_0$ $-$ 5s $^2$S & \citet{Ws66} \\
K I & 1.2522 & 4p $^2$P$_0$ $-$ 5s $^2$S & \citet{Ws66} \\
CH$_4$ & 1.3 - 1.5 & $\nu$$_2$ + 2$\nu$$_3$ & \citet{Da66} \\
H$_2$O & 1.33 - 1.43 & $\nu$$_2$ = 0, $\nu$$_1$ + $\nu$$_3$ = 2 & \citet{Au67} \\
H$_2$O & 1.43 - 1.52 & $\nu$$_2$ = 2, $\nu$$_1$ + $\nu$$_3$ = 1 & \citet{Au67} \\
CH$_4$ & 1.6 - 1.8 & 2$\nu$$_3$ & \citet{Da66} \\
H$_2$O & 1.71 - 1.80 & $\nu$$_2$ = -1, $\nu$$_1$ + $\nu$$_3$ = 2 & \citet{Au67} \\
H$_2$O & 1.80 - 2.08 & $\nu$$_2$ = 1, $\nu$$_1$ + $\nu$$_3$ = 1 & \citet{Au67} \\
CH$_4$ & 2.2 - 2.6 & $\nu$$_2$ + $\nu$$_3$ & \citet{Da66} \\
H$_2$ & centered at 2.4  & 1-0 Quadrupole (CIA) & \citet{Da66} \\
\enddata
\end{deluxetable}

\clearpage

\plotone{Burgasser.fig1.ps}

\plotone{Burgasser.fig2.ps}

\plotone{Burgasser.fig3.ps}

\end{document}